\documentclass[preprint,aps,showpacs]{revtex4}
\usepackage[normalem]{ulem}
\usepackage[dvips]{color}

\usepackage{mathrsfs}
\usepackage{graphicx}
\usepackage{amsmath}

\def\clap#1{\hbox to 0pt{\hss#1\hss}}

\usepackage[normalem]{ulem}  
\usepackage[dvips]{color} 

\renewcommand\sout{\bgroup \color{red} \ULdepth=-.5ex \ULset}

\begin{document}
\title{Energy release from hadron-quark phase transition in neutron stars and the axial $w$-mode of gravitational waves}
\author{Weikang Lin and Bao-An Li\footnote{Corresponding author: Bao-An\_Li$@$Tamu-Commerce.edu}}
\address{Department of Physics and Astronomy, Texas A$\&$M
University-Commerce, Commerce, Texas 75429-3011, USA}
\author{Jun Xu and Che Ming Ko}
\address{Cyclotron Institute and Department of Physics and Astronomy,
Texas A$\&$M University, College Station, Texas 77843-3366, USA}
\author{De-Hua Wen}
\address{Department of Physics, South China University of
Technology, Guangzhou 510641, P.R. China}

\date{\today}
\begin{abstract}
Describing the hyperonic and quark phases of neutron stars with an
isospin- and momentum-dependent effective interaction for the
baryon octet and the MIT bag model, respectively, and using the
Gibbs conditions to construct the mixed phase, we study the energy
release due to the hadron-quark phase transition. Moreover, the
frequency and damping time of the first axial $w$-mode of
gravitational waves are studied for both hyperonic and hybrid
stars. We find that the energy release is much more sensitive to
the bag constant than the density dependence of the nuclear
symmetry energy. Also, the frequency of the $w$-mode is found to
be significantly different with or without the hadron-quark phase
transition and depends strongly on the value of the bag constant.
Effects of the density dependence of the nuclear symmetry energy
become, however, important for large values of the bag constant
that lead to higher hadron-quark transition densities.
\end{abstract}

\maketitle

\section{Introduction}
Neutron stars (NSs) are among the most mysterious objects in the
Universe. With a typical mass of about $M=1.4M_{\odot}$ but a
radius of only about 12 km, the average density in NSs is several
times that in atomic nuclei. Their extreme compactness makes them
a natural laboratory to test our knowledge about general
relativity and properties of dense neutron-rich nuclear matter
\cite{Lat04}. Moreover, their generally large angular momentum and
possible quadrupole deformation make them strong candidates among
many possible sources to emit gravitational waves
(GWs)~\cite{Magg08}. It is well known that the estimated central
density of some NSs may be higher than the predicted hadron-quark
phase transition density, and thus there might be a quark core in
these so-called hybrid stars, see, e.g., ref.~\cite{Bombaci08} for
a recent review. Moreover, some originally hadronic NSs with
central densities below but close to the hadron-quark transition
density might increase their central densities due to, for
example, their spin-downs~\cite{Ara08} or the accretion of masses
from their binary companions~\cite{Chen98}. Therefore, the
hadron-quark phase transition may occur in their cores, turning
them into hybrid NSs as they evolve. Consequently, a
micro-collapse is expected to occur as a result of the softer
equation of state (EOS) of quark matter than that of hadronic
matter. Because of the difference in the binding energies for the
pure hyperonic and hybrid configurations of NSs, some energy will
be released from the NS after the hadron-quark phase transition.
To understand the mechanism of the hadron-quark phase transition
and the associated energy release, their dependence on properties
of the dense hadronic and quark matter, the way the energy release
being dissipated in NSs or carried away by gravitational waves,
and the gravitational wave signatures of the EOS of dense matter
and the expected hadron-quark phase transition are among the many
interesting questions currently under intense investigations in
neutron star physics and gravitational wave astronomy, see, e.g.,
refs.~\cite{Bombaci08,Hae89,Mar02,And09,Ara08b,Ben07,Pac10,Sch10}.

In a recent work involving some of us~\cite{Xu10}, several model
EOSs for hybrid stars were obtained from an isospin- and
momentum-dependent effective interaction (MDI)~\cite{Das03,Xu10}
for the baryon octet, the MIT bag model for the quark
matter~\cite{MIT1,MIT2} and the Gibbs construction for the
hadron-quark phase transition~\cite{Glen92,Glen01}. Since there
have been many other studies using similar approaches in the
literature, it is especially worth mentioning that while the
isospin-symmetric part of the hadronic EOSs used in this work is
constrained by the experimental data on collective flow and kaon
production in relativistic heavy-ion collisions \cite{Dan02}, the
symmetry energy term at sub-saturation densities is constrained in
a narrow range by the experimental data on isospin diffusion in
heavy-ion collisions at intermediate energies~\cite{LCK08}. The
obtained EOSs with various bag constants were then used in
studying the main features of the hadron-quark phase transitions
and the resulting mass-radius correlations of hybrid stars. With
this paper as a continuation of our recent work in
refs.~\cite{Listeiner,Krastev08a,Worley08,Xu10,Krastev08b,Wen09},
we make new contributions to the fast growing field of nuclear
astrophysics in two ways. First of all, critical for understanding
the physics of the hadron-quark phase transition are the EOSs of
both the hadronic and quark matter. Despite of the great efforts
made by both the astrophysics and the nuclear physics community,
our current knowledge about the EOS of dense matter either in the
hadronic or quark phase is still very poor. More specifically, for
the neutron-rich hadronic matter the most uncertain part of the
EOS is the density dependence of the nuclear symmetry energy which
encodes the energy associated with the neutron-proton
asymmetry~\cite{LCK08}. For the quark matter within the MIT bag
model, the most uncertain part of its EOS is the bag constant
which measures the inward pressure on the surface of the bag to
balances the outward pressure originating from the Fermi motion
and interactions of quarks confined in the bag. It is thus of
interest and also necessary to examine the relative effects of the
density dependence of the nuclear symmetry energy and the bag
constant on the total energy release due to the hadron-quark phase
transition in NSs. Secondly, a part of the energy release may be
carried away by GWs. The maximum amplitude of the latter can then
be obtained by assuming that all the released energy is available
for GWs. Besides testing a fundamental prediction of general
relativity, gravitational waves hold the great promises to open up
a completely new non-electromagnetic window into the Universe. On
the other hand, if detected in the future, the GWs may also help
probe properties of dense neutron-rich nuclear matter in NSs.
Thus, it is important to examine the imprints of the nuclear
symmetry energy and the bag constant on the frequency and damping
time of GWs. In this work, we focus on the first axial $w$-mode of
GWs due to the disturbance of the space-time
curvature~\cite{wmode}. We find that the energy release obtained
using the Gibbs construction for the quark-hadron phase transition
is much more sensitive to the bag constant than the density
dependence of the nuclear symmetry energy. Furthermore, the
frequency of the $w$-mode is found to be significantly different
with or without the hadron-quark phase transition and depends
strongly on the value of the bag constant $B$ for a given density
dependence of the nuclear symmetry energy. However, when a larger
bag constant is used such that the hadron-quark phase transition
happens at a higher baryon density, effects of the density
dependence of the nuclear symmetry energy also become very
important.

This article is organized as follows. In Section II, we summarize
the MDI interaction, the MIT bag model, and the resulting model
EOSs for hybrid stars obtained using the Gibbs construction. In
Section III, the total energy release from hadron-quark phase
transition in NSs and its dependence on the symmetry energy and
the bag constant are presented. In Section IV, the frequency and
damping time of the axial $w$-mode of GWs and their dependence on
the symmetry energy and the bag constant are studied. Finally, a
summary is given in Section V.

\section{Model equations of state for hybrid stars}

Since we are mostly interested in examining the relative effects
of the nuclear symmetry energy and the bag constant on the total
energy release during the hadron-quark phase transition in neutron
stars as well as the frequency and damping time of the $w$-mode of
emitted GWs, we first illustrate their effects on the EOS of
hybrid stars by using the model introduced in ref.\ \cite{Xu10}.
For completeness and to facilitate the discussions, we briefly
summarize the EOSs for hybrid stars obtained in ref. \cite{Xu10}
using the MDI interaction for the baryon octet, the MIT bag model
for the quark matter and the Gibbs construction for the
hadron-quark phase transition. We refer the reader to ref.\
\cite{Xu10} for details.
\begin{figure}[h]
\includegraphics[width=0.45\textwidth]{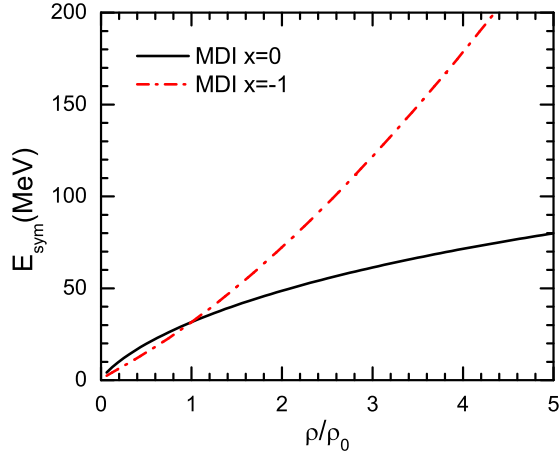}
\caption{(Color online) Density dependence of nuclear symmetry
energy from the MDI interaction with parameter $x=0$ and
$-1$.}\label{Esym}
\end{figure}

Assuming that the nucleon-hyperon and hyperon-hyperon interactions
have the same density and momentum dependence as the
nucleon-nucleon interaction with the interaction parameters fitted
to known experimental data at normal nuclear matter density, the
potential energy density of a hypernuclear matter due to
interactions between any two baryons is
\begin{eqnarray}
V_{bb^\prime} &=&\sum_{\tau_b,\tau^\prime_{b^\prime}}\left[
\frac{A_{bb^\prime}}{2 \rho_0} \rho_{\tau_b}
\rho_{\tau^\prime_{b^\prime}} + \frac{A^\prime_{bb^\prime}}{2
\rho_0} \tau_b \tau_{b^\prime} \rho_{\tau_b}
\rho_{\tau^\prime_{b^\prime}}\right. +
\left.\frac{B_{bb^\prime}}{\sigma+1}
\frac{\rho^{\sigma-1}}{\rho^\sigma_0}(\rho_{\tau_b}
\rho_{\tau^\prime_{b^\prime}} - x \tau_b \tau_{b^\prime}
\rho_{\tau_b} \rho_{\tau^\prime_{b^\prime}})\right. \\ \nonumber
&+ &\left.\frac{C_{{\tau_b},{\tau^\prime_{b^\prime}}}}{\rho_0}
\int
\int d^{3}pd^{3}p^{\prime }\frac{f_{\tau_b}(\vec{r},\vec{p}%
)f_{\tau^\prime_{b^\prime}}(\vec{r},\vec{p}^{\prime
})}{1+(\vec{p}-\vec{p}^{\prime })^{2}/\Lambda ^{2}}\right],
\label{Vbb}
\end{eqnarray}
where $b$ ($b^\prime$) denotes the baryon octet included in the
present study, i.e., $N$, $\Lambda$, $\Sigma$, and $\Xi$. The
conventions for the isospin are $\tau_N=-1$ for neutron and $1$
for proton, $\tau_\Lambda=0$ for $\Lambda$, $\tau_\Sigma=-1$ for
$\Sigma^-$, $0$ for $\Sigma^0$ and $1$ for $\Sigma^+$, and
$\tau_\Xi=-1$ for $\Xi^-$ and $1$ for $\Xi^0$. The total baryon
density is given by $\rho = \sum_b \sum_{\tau_b} \rho_{\tau_b}$,
and $f_{\tau_b}({\vec r},{\vec p})$ is the phase-space
distribution function of particle species $\tau_b$. For hyperons
in symmetric nuclear matter at saturation density, their
potentials are fixed at $U^{(N)}_\Lambda(\rho_N^{} = \rho_0^{}) =
- 30 ~\text{MeV}$, $U^{(N)}_\Xi(\rho_N^{} = \rho_0^{}) = - 18
~\text{MeV}$ and $U^{(N)}_\Sigma(\rho_N^{} = \rho_0^{}) = 30
~\text{MeV}$ for $\Lambda, \Xi$ and $\Sigma$, respectively. The
parameter $x$ is used to model the isospin dependence of the
interaction between two baryons. Its value can be adjusted to
mimic different density dependence of the nuclear symmetry energy
and is taken to be $0$ or $-1$ in the present study. Shown in
Fig.\ \ref{Esym} is the corresponding density dependence of the
nuclear symmetry energy $E_{\rm sym}(\rho)$. It is worth noting
that the latest experimental constraints on $E_{\rm sym}(\rho)$ at
sub-saturation densities are consistent with but span a region
larger than the one covered by the $x=0$ and $x=-1$ curves
\cite{cxu10c}. The experimental constraints on $E_{\rm sym}(\rho)$
at supra-saturation densities are, on the other hand, still very
uncertain. In fact, the high density behavior of $E_{\rm
sym}(\rho)$ is considered the most uncertain part of the EOS of
dense neutron-rich nuclear matter \cite{kut}. One of the purposes
of this work is to study effects of the symmetry energy. As shown
in Fig.\ \ref{Esym}, the choice of $x=0$ and $x=-1$ allows the
variation of the symmetry energy in a large range at
supra-saturation densities.

\begin{figure}[h]
\centering
\includegraphics[height=5.5cm]{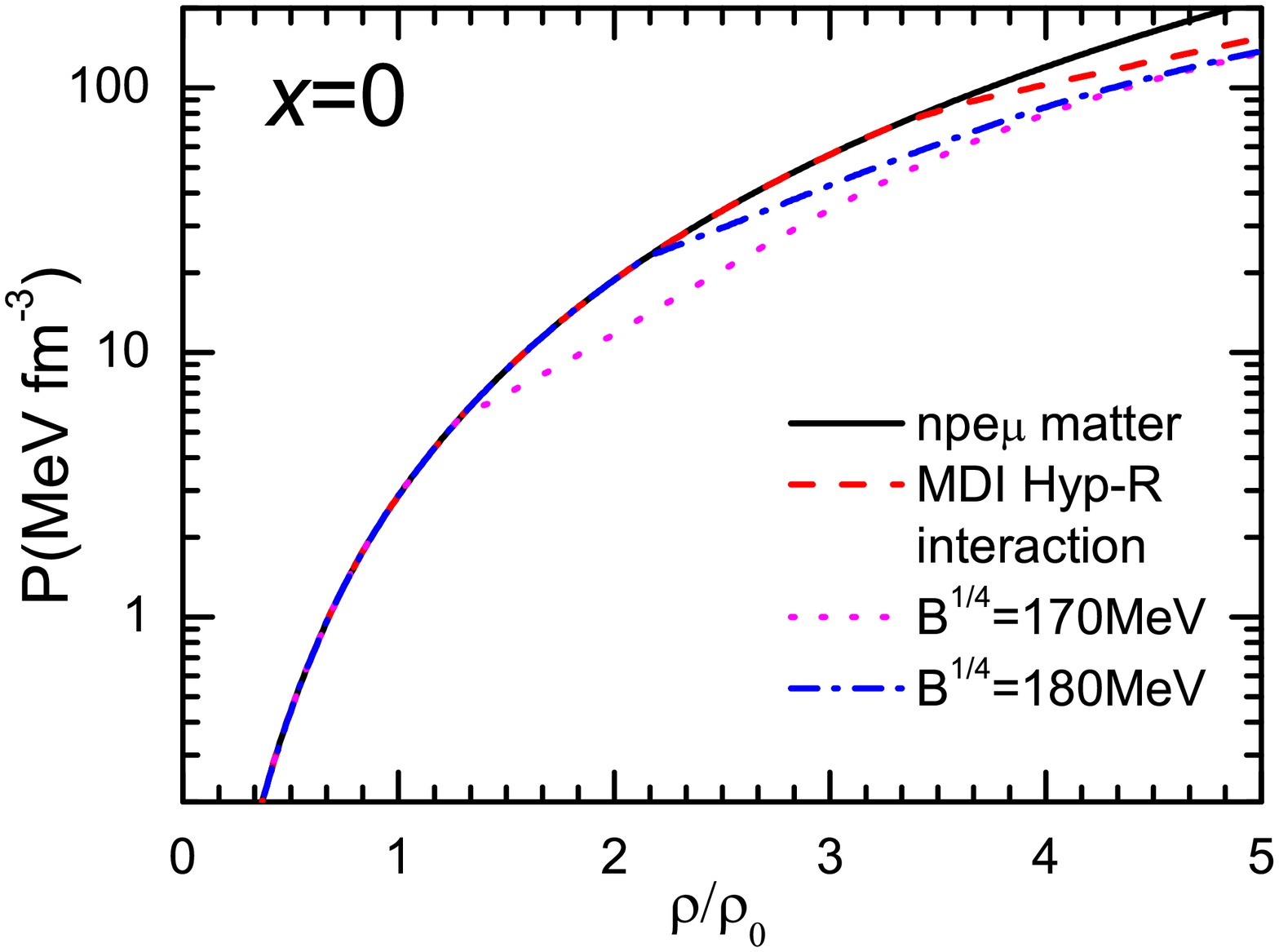}
\includegraphics[height=5.5cm]{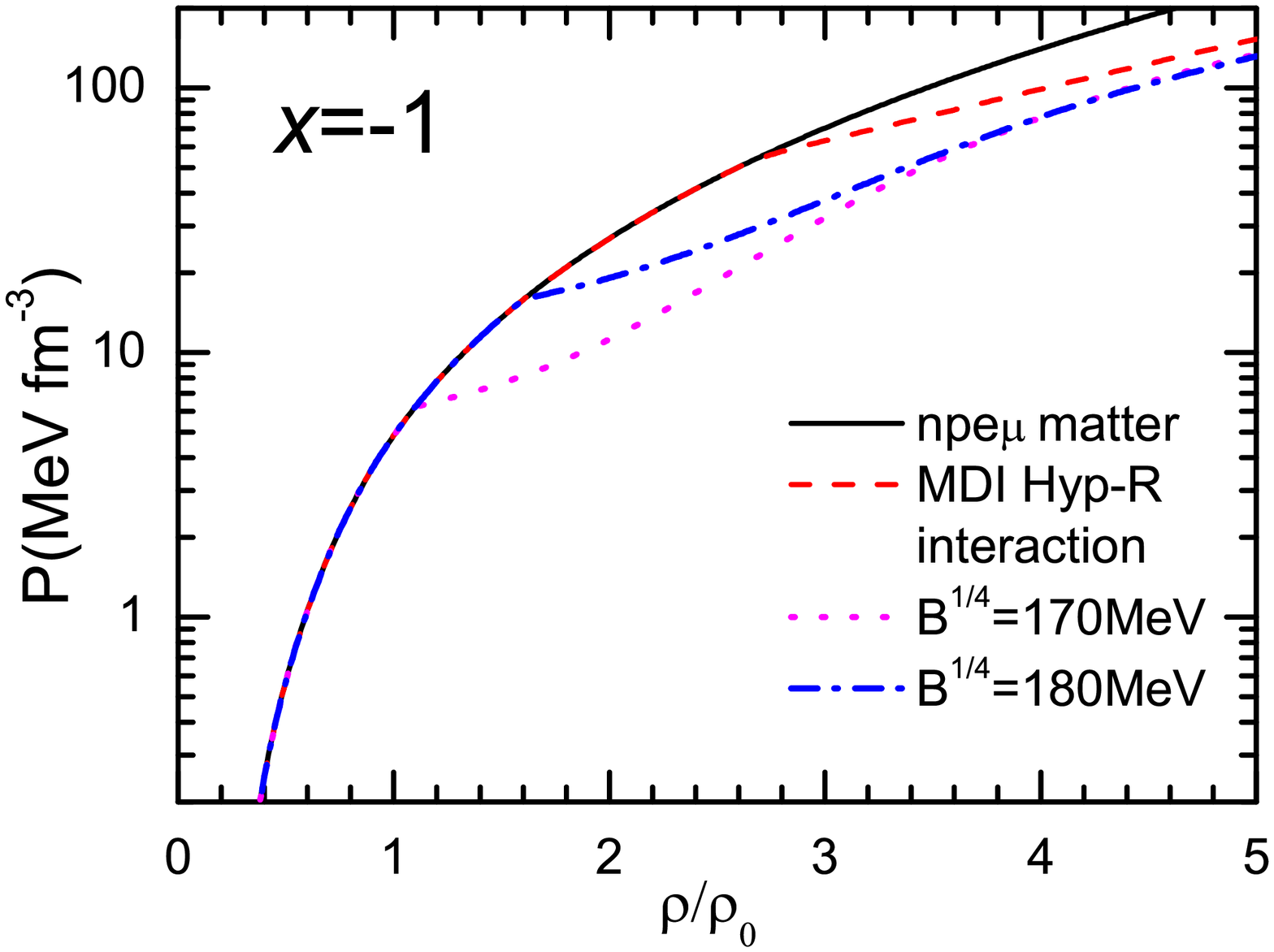}
\caption{(Color online) The EOSs of pure $npe\mu$ matter
(nucleonic), hyperonic matter (MDI Hyp-R interaction) and hybrid
stars with $B^{1/4}=180$ MeV and $170$ MeV, and symmetry energy
parameter $x=0$ (left window) and $x=-1$ (right
window).}\label{Eosx}
\end{figure}

To obtain the complete EOS for hybrid stars, besides the EOS for
hyperons described above, the MIT bag model for the quark matter
\cite{MIT1,MIT2} and the Gibbs construction for the hadron-quark
phase transition \cite{Glen92,Glen01} have been used in
ref.~\cite{Xu10}. As in previous studies, see, e.g., refs.\
\cite{XCLM09a,XCLM09b}, the hybrid star is divided into three
parts from the center to the surface: the liquid core, the inner
crust, and the outer crust. It is in the liquid core where the
matter can be pure hadrons, quarks or a mixture of the two. In the
inner crust, a parameterized EOS of $P=a+b \epsilon^{4/3}$ is used
as in refs.\ \cite{XCLM09a,XCLM09b}. The outer crust usually
consists of heavy nuclei and the electron gas, where the BPS
EOS~\cite{BPS} is used. The transition density $\rho_{\rm t}$
between the liquid core and the inner crust has been consistently
determined in refs.~\cite{XCLM09a,XCLM09b}. Taking the density
separating the inner from the outer crust to be $\rho_{\rm
out}=2.46\times10^{-4}$ fm$^{-3}$, the parameters $a$ and $b$ can
then be determined using the pressures and energy densities at
$\rho_{\rm t}$ and $\rho_{\rm out}$. Shown in Fig.~\ref{Eosx} are
the EOSs with $x=0$ and $x=-1$, respectively, both using the MIT
bag constant $B^{1/4}=180$ MeV and 170 MeV. For comparisons, the
pure $npe\mu$ (labeled as nucleonic) and hyperonic (labeled as MDI
Hyp-R interaction) EOSs are also included. As it is well known,
the appearance of hyperons and the hadron-quark phase transition
softens the EOS of neutron star matter. Also, it is worth noting
that the adiabatic coefficient $\gamma=d\log(P)/d\log(\rho)$ at
saturation density is 2.63 and 2.57 for $x=0$ and $x=-1$,
respectively. The EOSs with $x=0$ and $x=-1$ are thus about the
same below and around the saturation density. However, it is
interesting to see that the starting point and the degree of
softening due to the appearance of hyperons are sensitive to
$E_{\rm sym}(\rho)$ at high densities. Moreover, the $E_{\rm
sym}(\rho)$ also affects appreciably the starting point of the
hadron-quark mixed phase especially with the larger bag constant.
Nevertheless, it is obvious that the starting point is much more
sensitive to the bag constant $B$ for a given symmetry energy
parameter $x$. These features are consistent with those first
noticed by Kutschera et al. \cite{Kut00}. It is also important to
emphasize that since the mixed phase is described using the Gibbs
instead of the Maxwell construction, the energy density thus
increases continuously with pressure across the mixed phase
without a jump \cite{Bhatt09}. Consequently, the Sidov criterium
that a stable hybrid star must satisfy \cite{Mar02,Sei71}, i.e.,
$\frac{\epsilon_q}{\epsilon_H}<\frac{3}{2}(1+\frac{P}{\epsilon_H})$
with P, $\epsilon_q$ and $\epsilon_H$ denoting, respectively, the
pressure, the energy density of quarks and hadrons at the
transition point, is automatically satisfied.

\section{Total energy release due to hadron-quark phase transition in neutron stars}

The amount of energy release during the micro-collapse of a
neutron star triggered by the hadron-quark phase transition is
given by the change in its binding energy before and after the
phase transition. The binding energy of a neutron star can be
obtained from first solving the Tolman-Oppenheimer-Volkoff (TOV)
equation with the corresponding EOS ($G=c=1$),
\begin{eqnarray}\label{TOV}
&&\frac{dP(r)}{dr}=-\frac{(m(r)+4\pi
r^3P(r))(\epsilon(r)+P(r))}{r(r-2m(r))},\\\nonumber
&&\frac{dm(r)}{dr}=4\pi \epsilon(r) r^2,
\end{eqnarray}
where $P(r)$ and $\epsilon(r)$ are the pressure and the energy
density at radius $r$. The binding energy of a neutron star is
then given by \cite{Weinberg}
\begin{equation}\label{eq:E_G}
E_b=M_g-M_{\rm bar},
\end{equation}
where $M_g$ is the gravitational mass of the neutron star measured
from infinity \cite{Gravity}, i.e.,
\begin{equation}
M_g=\int_0^R4\pi r^2\epsilon(r)dr.
\end{equation}
The $M_{\rm bar}$ is the baryonic mass of the neutron star,
namely, the mass when all the matter in the neutron star is
dispersed to infinity \cite{Weinberg}. It can be calculated from
$M_{\rm bar}=NM_B$, where $M_B=1.66\times10^{-24}g$ is the nucleon
mass and $N$ is the total number of baryons \cite{Weinberg}, i.e.,
\begin{equation}
N=\int^R_04\pi
r^2\left[1-\frac{2m(r)}{r}\right]^{-1/2}\rho_B(r)dr\,
\end{equation}
with $\rho_B(r)$ being the baryon density profile of the neutron
star. While the total energy release $E_g$ during the phase
transition is the difference of Eq. \eqref{eq:E_G} before and
after the micro-collapse, it reduces to the difference in
gravitational mass for a hadronic (h) and a hybrid star (q) as a
result of baryon number conservation, namely,
\begin{equation}
E_g=M_{g,h}-M_{g,q}.
\end{equation}
\begin{figure}[h]
\includegraphics[width=0.55\textwidth]{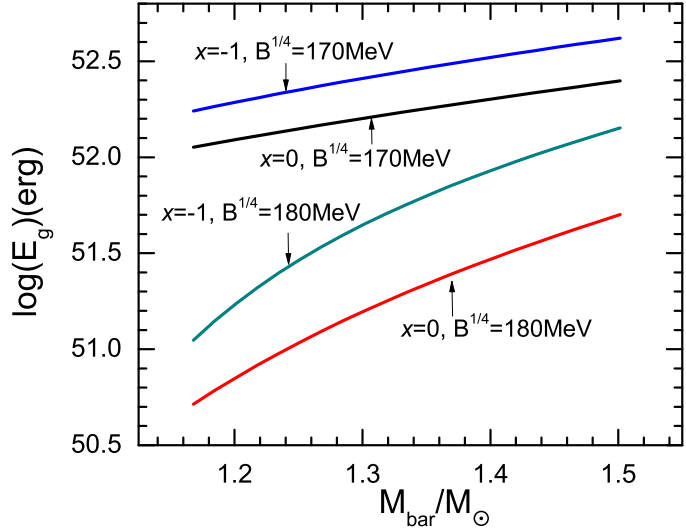}
\caption{Total energy release due to the hadron-quark phase
transition as a function of the baryonic mass of a neutron
star.}\label{Ere}
\end{figure}

\begin{figure}[h]
\includegraphics[height=5.5cm,bb = 0 0 570 430]{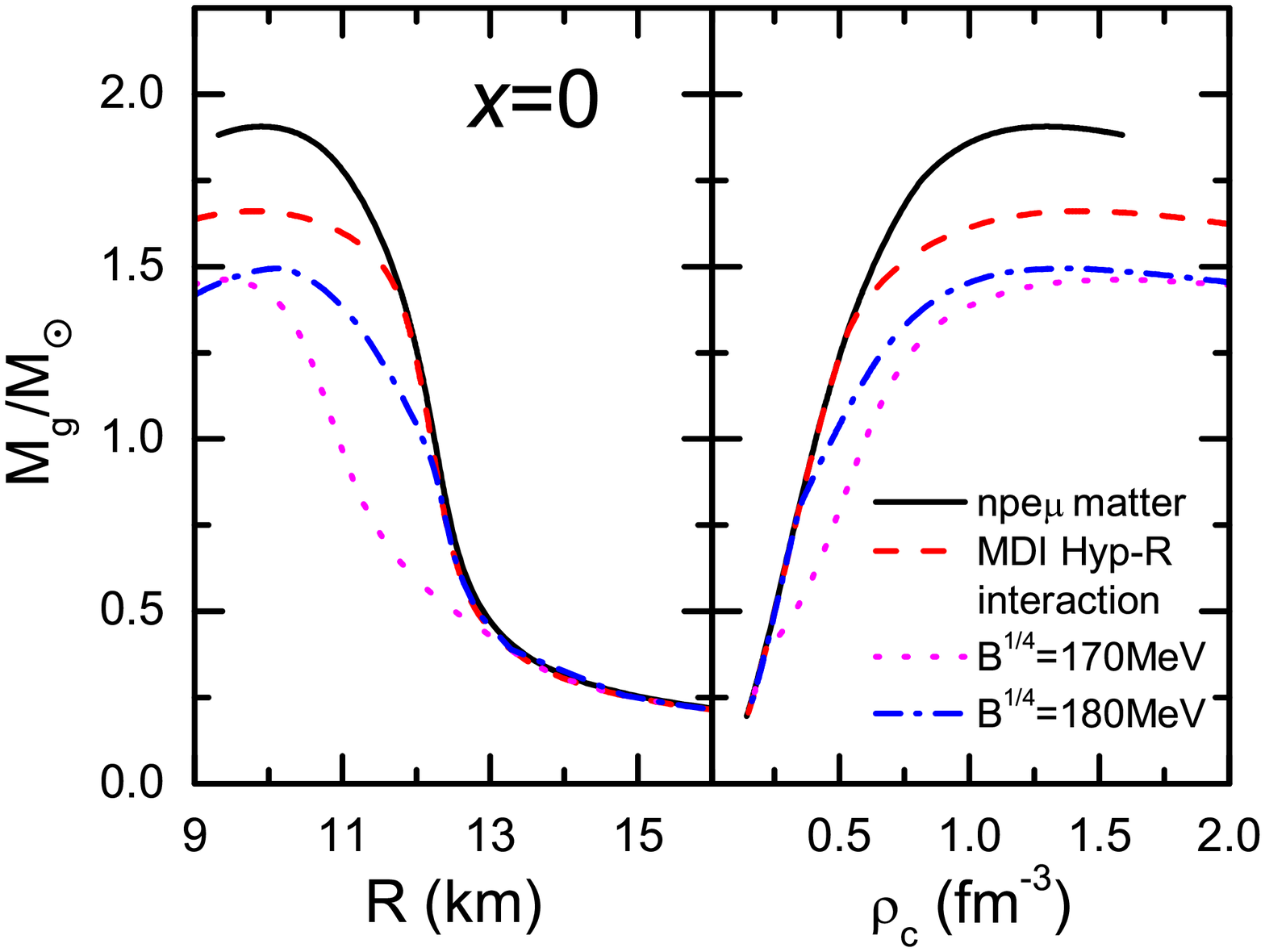}
\includegraphics[height=5.5cm]{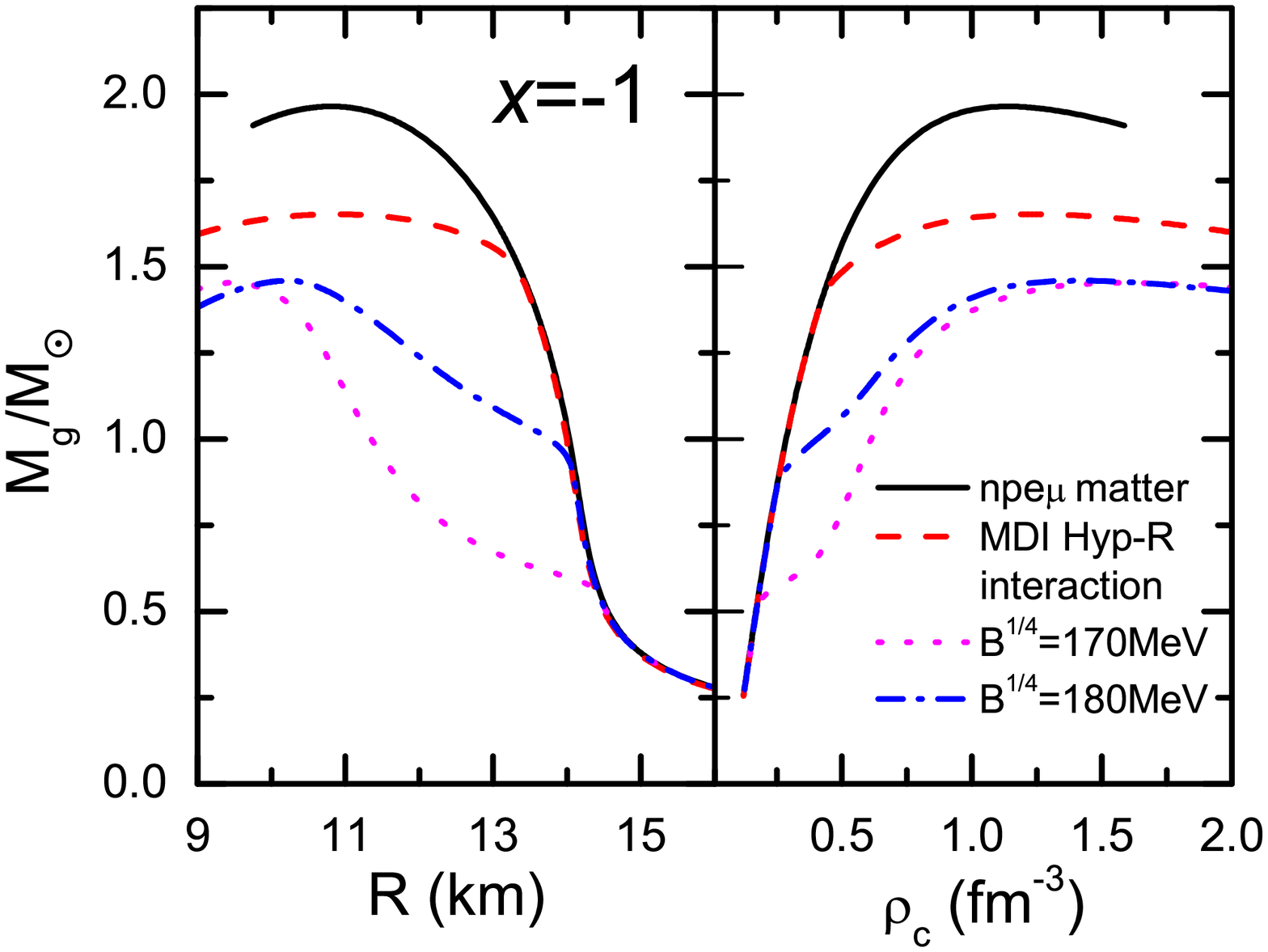}
\caption{(Color online) The mass-radius and mass-central density
correlations for pure $npe\mu$ matter (nucleonic), hyperonic
matter (MDI Hyp-R interaction) and hybrid stars with $B^{1/4}=180$
MeV and $170$ MeV, and symmetry energy parameter $x=0$ (left
window) and $x=-1$ (right window).}\label{MRx}
\end{figure}

\begin{table}[h]
\caption{The baryon number ($N$), gravitational masses of
hyperonic stars ($M_g(H)$), hybrid stars ($M_g(HQ)$), the quark
core including the mixed phase ($M_g(\text{core})$), the radii of
hyperonic ($R(H)$) and hybrid stars ($R(HQ))$ as well as the
energy release ($E_g$) with $x=0$ and $B^{1/4}=170~{\rm MeV}$. All
masses are in unit of $M_{\odot}$ and radii in km.}\label{tab:1}
\begin{tabular}{|l|l|l|l|l|l|l|l|c|}
\hline $N$ & $M_{g}(H)$ & $M_{g}(HQ)$ & $M_{g}(\text{core})$ & $R(H)$ & $R(HQ)$ & $\log(E_g)(erg)$\\ \hline\hline%
1.4E57 & 1.0861 & 1.0798 & 0.886659 & 11.2917 & 10.1679 & 52.0517 \\ \hline%
1.5E57 & 1.1570 & 1.1491 & 0.971092 & 11.3044 & 10.1256 & 52.1492 \\ \hline%
1.6E57 & 1.2270 & 1.2173 & 1.054172 & 11.3041 & 10.0691 & 52.2379 \\ \hline%
1.7E57 & 1.2960 & 1.2843 & 1.135890 & 11.2909 & 9.9935 & 52.3199 \\ \hline%
1.8E57 & 1.3641 & 1.3501 & 1.217399 & 11.2620 & 9.8796 & 52.3977 \\ \hline%
\end{tabular}
\end{table}
\begin{table}[h]
\caption{Same as \ref{tab:1} but with $x=0$ and $B^{1/4}=180$ MeV.
}\label{tab:2}
\begin{tabular}{|l|l|l|l|l|l|l|l|c|}
\hline $N$ & $M_{g}(H)$ & $M_{g}(HQ)$ & $M_{g}(\text{core})$ & $R(H)$ & $R(HQ)$ & $\log(E_g)(erg)$ \\ \hline\hline%
1.4E57 & 1.0861 & 1.0858 & 0.335018 & 11.2917 & 11.0960 & 50.7137 \\\hline%
1.5E57 & 1.1570 & 1.1564 & 0.457409 & 11.3044 & 11.0076 & 51.0370 \\\hline%
1.6E57 & 1.2270 & 1.2259 & 0.582411 & 11.3041 & 10.8971 & 51.2965 \\\hline%
1.7E57 & 1.2960 & 1.2942 & 0.708860 & 11.2909 & 10.7625 & 51.5124 \\\hline%
1.8E57 & 1.3641 & 1.3612 & 0.838033 & 11.2620 & 10.5902 & 51.7018 \\\hline%
\end{tabular}
\end{table}
\begin{table}[h]
\caption{Same as \ref{tab:1} but with $x=-1$ and $B^{1/4}=170$ MeV.
}\label{tab:3}
\begin{tabular}{|l|l|l|l|l|l|l|l|c|}
\hline $N$ & $M_{g}(H)$ & $M_{g}(HQ)$ & $M_{g}(\text{core})$ & $R(H)$ & $R(HQ)$ & $ \log(E_g)(erg)$ \\ \hline\hline%
1.4E57 & 1.0911 & 1.0813 & 0.865139 & 12.5570 & 10.3024 & 52.2408 \\ \hline%
1.5E57 & 1.1632 & 1.1507 & 0.953706 & 12.5975 & 10.2280 & 52.3516 \\ \hline%
1.6E57 & 1.2346 & 1.2189 & 1.040034 & 12.6178 & 10.1446 & 52.4498 \\ \hline%
1.7E57 & 1.3052 & 1.2859 & 1.124924 & 12.6185 & 10.0418 & 52.5385 \\ \hline%
1.8E57 & 1.3749 & 1.3516 & 1.209269 & 12.5998 & 9.8992 & 52.6205 \\ \hline%
\end{tabular}
\end{table}
\begin{table}[h]
\caption{Same as \ref{tab:1} but with $x=-1$ and $B^{1/4}=180$ MeV.
}\label{tab:4}
\begin{tabular}{|l|l|l|l|l|l|l|l|c|}
\hline $N$ & $M_{g}(H)$ & $M_{g}(HQ)$ & $M_{g}(\text{core})$ & $R(H)$ & $R(HQ)$ & $ \log(E_g)(erg)$ \\ \hline\hline%
1.4E57 & 1.0911 & 1.0904 & 0.389260 & 12.5570 & 11.8321 & 51.0467 \\ \hline%
1.5E57 & 1.1632 & 1.1616 & 0.563910 & 12.5975 & 11.4949 & 51.4673 \\ \hline%
1.6E57 & 1.2346 & 1.2314 & 0.716042 & 12.6178 & 11.2027 & 51.7548 \\ \hline%
1.7E57 & 1.3052 & 1.3999 & 0.855991 & 12.6185 & 10.9207 & 51.9731 \\ \hline%
1.8E57 & 1.3749 & 1.3670 & 0.990235 & 12.5998 & 10.6140 & 52.1525 \\ \hline%
\end{tabular}
\end{table}
Shown in Fig.\ \ref{Ere} is the energy release as a function of
the baryonic mass $M_{\rm bar}/M_{\odot}$ of a neutron star. To be
more quantitative, shown in tables \ref{tab:1}, \ref{tab:2},
\ref{tab:3} and \ref{tab:4} are detailed comparisons of the baryon
number ($N$), gravitational masses of hyperonic stars ($M_g(H)$),
hybrid stars ($M_g(HQ)$), the quark core including the mixed phase
($M_g(\text{core})$), the radii of hyperonic ($R(H)$) and hybrid
stars ($R(HQ)$) as well as the energy release ($E_g$). It is seen
that the energy release increases with $M_{\rm bar}/M_{\odot}$ and
is higher with the smaller ($B^{1/4}=170$ MeV) bag constant $B$
but stiffer ($x=-1$) symmetry energy. Effects of varying the bag
constant $B$ are obviously more significant than varying the
symmetry energy parameter $x$, especially on the core mass and
thus the energy release. The variation of the bag constant $B$
also affects appreciably the radii of hybrid stars. On the other
hand, the variation of the symmetry energy parameter $x$ only has
appreciable effects on the radii of both hyperonic and hybrid
stars. It's effects on the energy release is much smaller than the
bag constant $B$. It is worth noting that it was first shown in
ref. \cite{Newton} that the binding energies of NSs consisting of
pure $npe\mu$ matter depend strongly on the density dependence of
the nuclear symmetry energy. It is easily understandable that the
difference in binding energies before and after the hadron-quark
phase transition becomes less sensitive to the symmetry energy.
Some of these features can be better understood from inspecting
the gravitational mass-radius and mass-central density
correlations with and without the hadron-quark phase transition
using different bag constants and symmetry energy functionals.
Shown in Fig.\ \ref{MRx} are the gravitational mass $M_g$ as a
function of radius $R$ and central density $\rho_c$ for several
relevant cases. It is seen that with $B^{1/4}=170$ MeV the phase
transition happens at a significantly lower density, resulting in
a smaller maximum mass than the case with $B^{1/4}=180$ MeV. On
the other hand, for a given bag constant $B$, the hyperonic EOS is
stiffer with a stiffer symmetry energy, leading to a larger radius
and thus a lower central density and a smaller maximum mass. The
mass difference before and after the hadron-quark phase transition
is therefore also larger.

To see how the total energy release is related to emitted GWs, we
add a few comments as follows. They are also useful for explaining
why we study in detail in the next section the frequency and
damping time of the $w$-mode of GWs. Generally, the magnitude of
GWs is denoted by the gravitational strain amplitude in the
following wave form:
\begin{equation}
h(t)=h_0e^{-(t/\tau_{g\omega})}\sin(\omega_0 t),
\end{equation}
where $h_0$ is the initial amplitude, $\omega_0$ is the angular
frequency and $\tau_{g\omega}$ is the damping timescale. Assuming
all energy release $E_g$ is available to GW emission, the maximum
value of the initial amplitude is then \cite{Amp}
\begin{equation}
h_0=\frac{4}{\omega_0
D}\left[\frac{E_g}{\tau_{g\omega}}\right]^{1/2},
\end{equation}
where $D$ is the distance of the source from the detector. Thus,
the maximum strain amplitude is directly determined by the energy
release. However, not all the energy release is available for
emitting gravitational waves~\cite{Mar02} as the energy can be
dissipated by other mechanisms. In fact, most of the energy
release from the hadron-quark phase transition would be used to
excite radial oscillations \cite{Sot02}. Moreover, the pure radial
models do not generate any gravitational waves unless they are
coupled with rotation \cite{Chau67}. However, because of angular
momentum conservation, rotating neutron stars are more likely to
be formed during supernova explosions. If there are some other
dissipating mechanisms of timescale $\tau_d$ during the
hadron-quark phase transition in neutron stars, the fraction of
energy dissipated by gravitational waves is then approximately
$f_{g}=1/(1+\tau_{g\omega}/\tau_d)$~\cite{Mar02}. Thus, the
relative damping timescale of the GWs with respect to that due to
other mechanisms determines the fraction of the total energy
release that can be carried away by GWs. In the literature, time
scales of various GW modes and various energy dissipation
mechanisms have been considered, see, e.g., refs.
\cite{Mar02,Chau67} and references therein. As for many other
interesting questions in the field, the conclusions so far are
still very model dependent.

\section{Frequency and damping time of the axial $w$-mode of gravitational waves}

GWs from non-radial oscillations of NSs carry important
information about their internal
structures~\cite{Kok92,Nil98,Ben05}. One special type of normal
modes of oscillations that only exists in general relativity is
the so called $w$-mode associated with the perturbation of the
space-time curvature and for which the motion of the fluid is
negligible \cite{wmode}. While the frequencies of the $w$-mode is
significantly above the operating frequencies of existing GW
detectors, it is nevertheless useful to further investigate what
information about the internal structure of NSs may be revealed
from a future detection of the $w$-mode of GWs.

According to Chandrasekhar \& Ferrari \cite{wmode}, the axial
perturbation equations for a static neutron star can be simplified
by introducing a function $z(r)$ that is constructed from the
radial part of the perturbed axial metric components.  It
satisfies a Schr\"odinger-like differential equation
\begin{equation}\label{RW}
\frac{d^{2}z}{dr_{*}^{2}}+[\omega^{2}-V(r)]z=0
\end{equation}
where $\omega (=\omega_{0}+i\omega_{i})$  is the complex
eigen-frequency of the axial $w$-mode and $r_{*}$ is the tortoise
coordinate defined by
\begin{equation}\label{rstar}
 \frac{d}{dr_{*}}=e^{\lambda-\nu}\frac{d}{dr}
\end{equation}
where the $e^\nu$ and $e^\lambda$ are the metric functions given
by the line element for a static neutron star \cite{wmode}. Inside
a neutron star, the potential function $V$ is defined by
\begin{equation}\label{VP}
V=\frac{e^{2\nu}}{r^{3}}[l(l+1)r+4\pi r^{3}(\rho(r)-P(r))-6m(r)]
\end{equation}
with $l$ the spherical harmonics index (used in describing the
perturbed metric and only the case $l=2$ is considered here),
$\rho(r)$ and $P(r)$  the density and pressure, and $m(r)$ the
mass inside radius $r$, respectively. Outside the neutron star,
Eq.~(\ref{VP}) reduces to
\begin{equation}
V=\frac{r-2M_g}{r^{4}}[l(l+1)r-6M_g]
\end{equation}
where $M_g$ is the total gravitational mass of the neutron star.
The solutions to this problem are subject to a set of boundary
conditions (BC) constructed by Chandrasekhar \& Ferrari
\cite{wmode}: regular BC at the neutron star center, continuous BC
at the surface and behaving as a purely outgoing wave at infinity.
In a recent work \cite{Wen09} using the continued fraction method
\cite{Lei93,Ben99}, some of us have studied the frequency and
damping time of the axial $w$-mode for NSs containing only the
$npe\mu$ matter. It was found that the density dependence of the
symmetry energy has strong imprints on both the frequency and
damping time of the axial $w$-mode. In this section, we examine
how the appearance of hyperons and the hadron-quark phase
transition may affect the frequency and damping time of the axial
$w$-mode. As mentioned earlier, we focus on the relative effects
of the symmetry energy and the bag constant on the first $w$-mode.

\begin{figure}[h]
\label{freq}
\includegraphics[height=5.5cm]{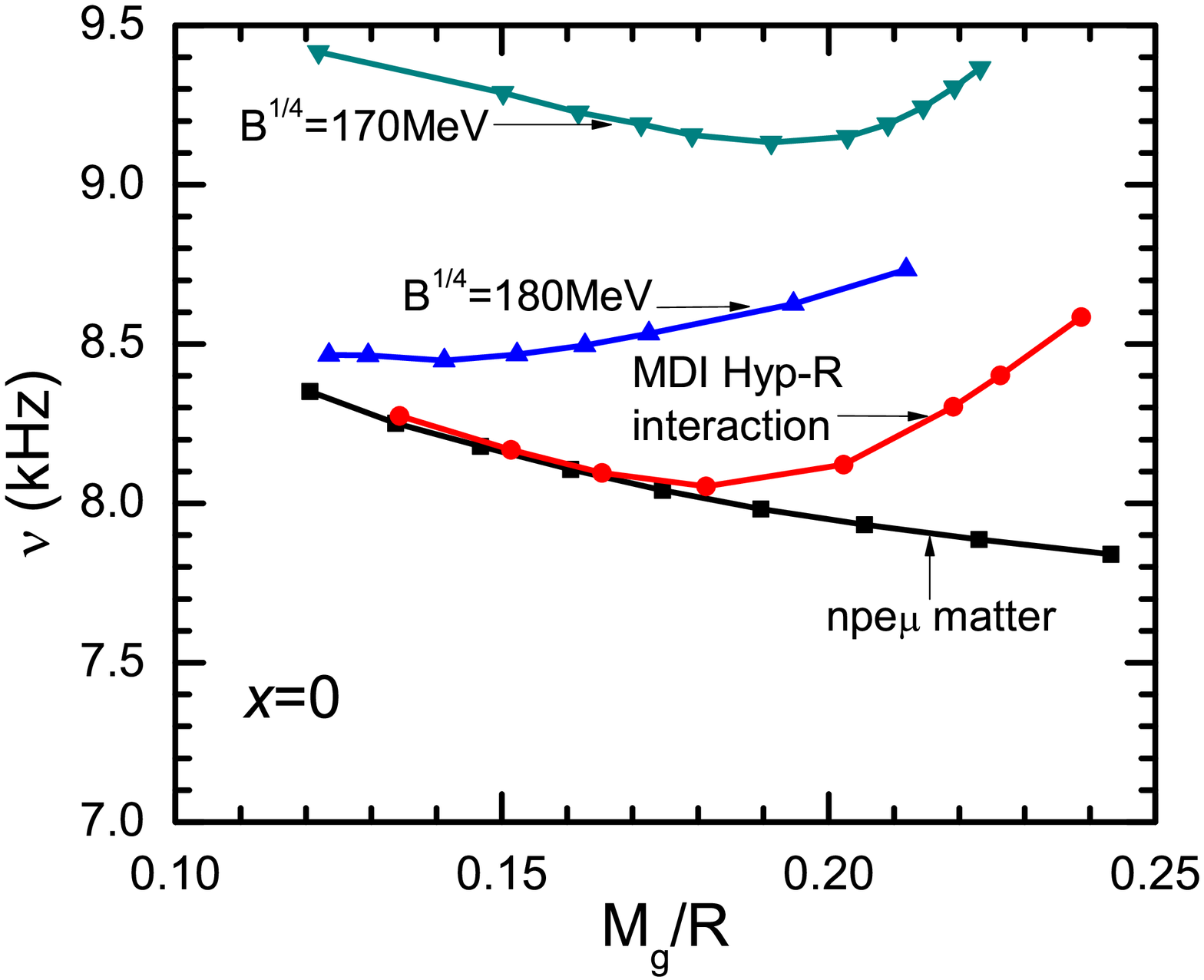}
\includegraphics[height=5.5cm]{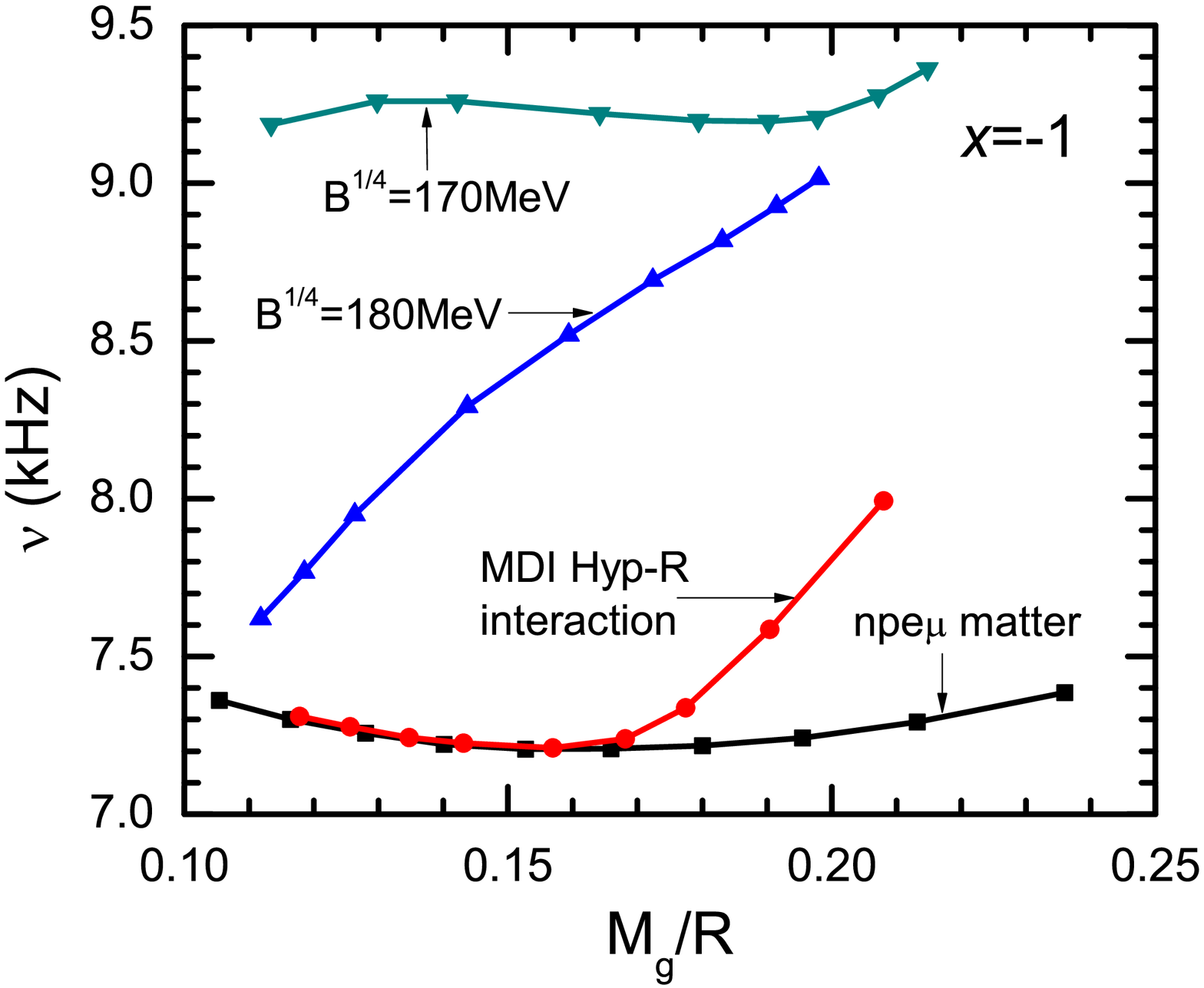}
\caption{(Color online) Frequency of the first $w$-mode as a
function of the neutron star compactness $M_g/R$ with the symmetry
energy parameter $x=0$ (left window) and $x=-1$ (right window).}
\end{figure}

Shown in Fig. 5 \ is the frequency as a function of the neutron
star compactness $M_g/R$ for the four situations considered, i.e.,
pure $npe\mu$ star, hyperonic star, and hybrid stars with two
different values for the bag constant. It is interesting to see
that the frequency is very sensitive to the EOS used. This is
consistent with the earlier findings in refs. \cite{Ben05,Wen09}
where a large ensemble of hadronic EOSs were used to describe the
$npe\mu$ matter. Compared to these previous studies, it is seen
that as the EOS softens when hyperons appear or the hadron-quark
phase transition happens, the frequency increases quickly. It is
also seen that the effect of the hadron-quark phase transition is
dramatic. Comparing results in the two figures with $x=0$ and
$x=-1$, it is seen that the bag constant $B$ plays a much stronger
role than the symmetry energy parameter $x$.

\begin{figure}[h]
\label{damt}
\includegraphics[height=5.5cm]{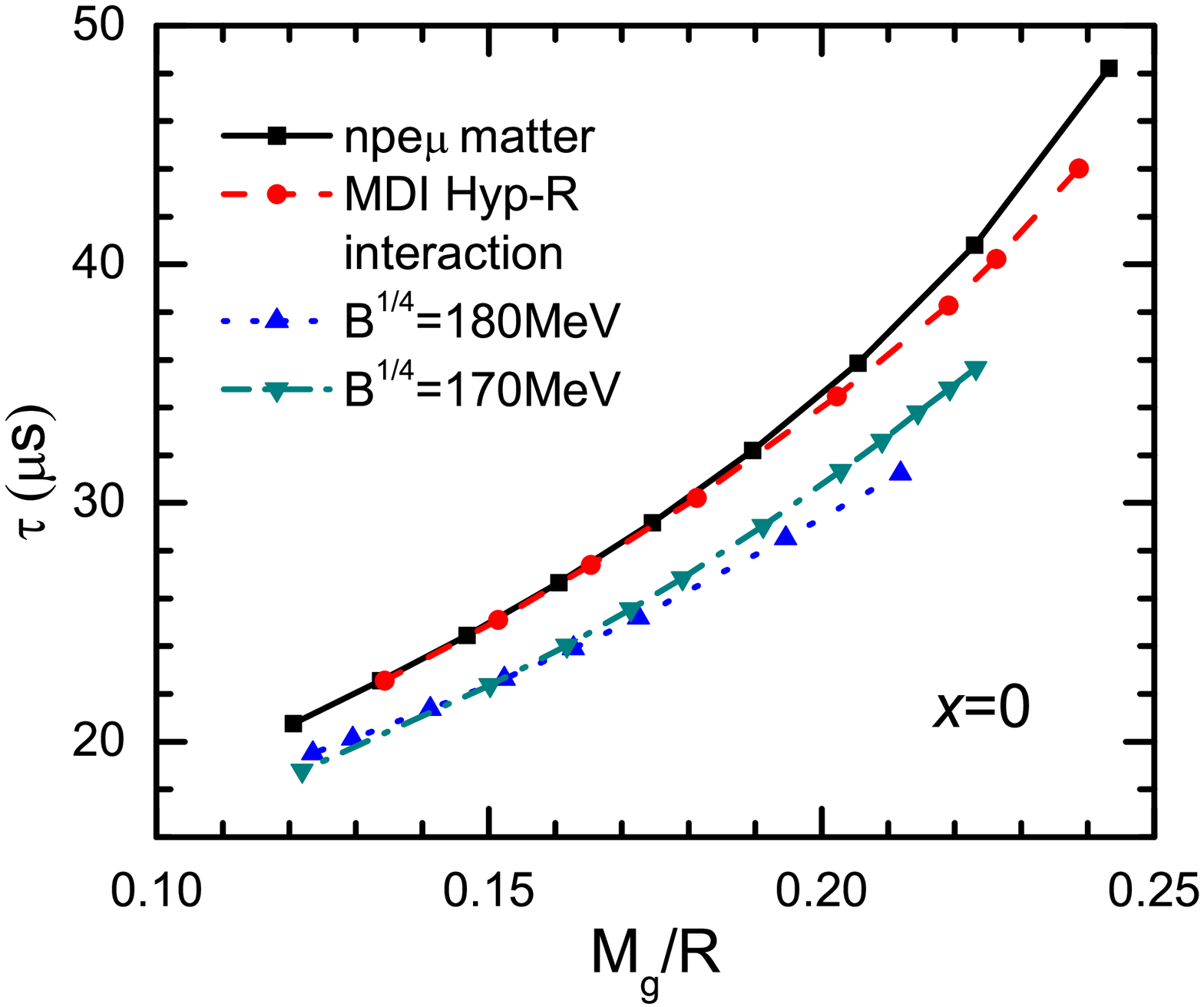}
\includegraphics[height=5.5cm]{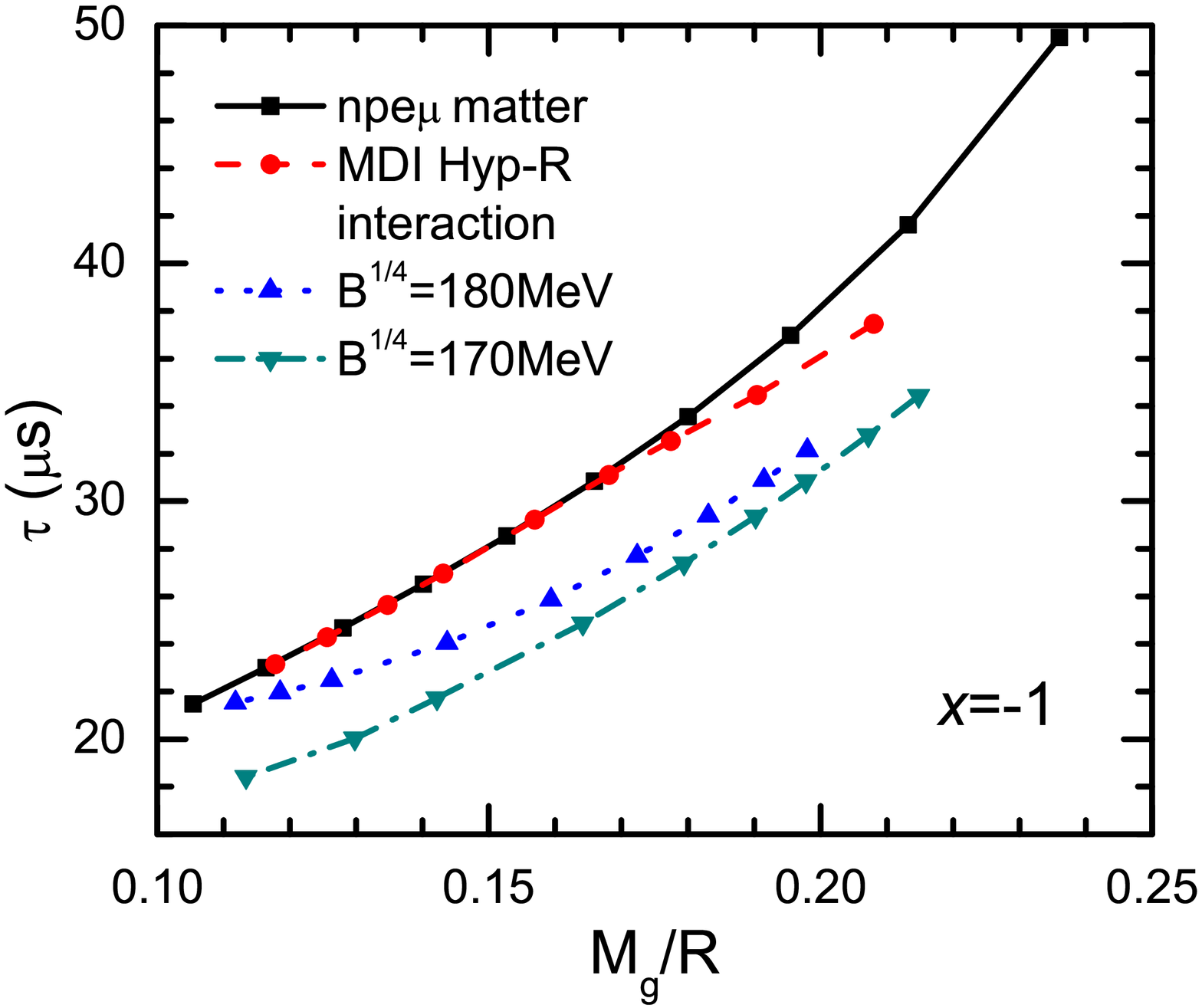}
\caption{(Color online) Damping time of the first $w$-mode as a
function of the neutron star compactness $M_g/R$ with the symmetry
energy parameter $x=0$ (left window) and $x=-1$ (right window).}
\end{figure}

Shown in Fig. 6 is the damping time as a function of $M_g/R$ for
the four situations considered. It is seen that the damping time
is longer for more compact NSs. Effects of the EOS on the damping
time are appreciable but not as strong as those on the frequency.
This might turn out to be an advantage for extracting information
about the EOS of neutron star matter from analyzing the GW
signals.

\begin{figure}[h]
\label{tvcor}
\includegraphics[width=0.55\textwidth]{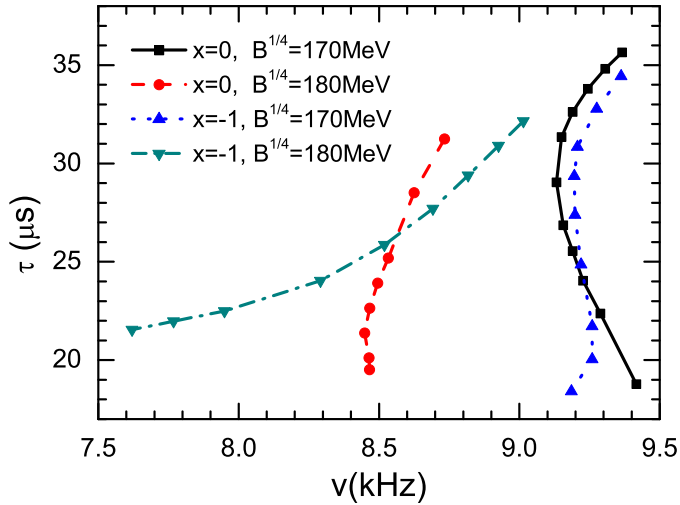}
\caption{(Color online) Damping time versus frequency of the first
$w$-mode for hybrid stars.}
\end{figure}

To assess more clearly the relative effects of the symmetry energy
and bag constant, we present in Fig. 7 the correlation between the
damping time and the frequency for hybrid stars. For the softer
EOS with $B^{1/4}=170$ MeV, the effect of the symmetry energy is
small. On the other hand, for the stiffer EOS with $B^{1/4}=180$
MeV, the density dependence of the symmetry energy can affect the
frequency significantly. This is because the hadron-quark
transition happens at a higher baryon density when a large bag
constant is used as shown in Fig. \ref{Eosx}. The symmetry energy
in the hadronic phase then also plays an important role in
determining the structure of NSs. Nevertheless, comparing
frequencies with the same $x$ parameter but different values of
$B$, it is clear that the bag constant has a much stronger effect
as it changes the underlying EOS of dense matter more
significantly.

\section{Summary}
Using an isospin- and momentum-dependent effective interaction for
the baryon octet and the MIT bag model to describe, respectively,
the hadronic and quark phases of neutron stars, we have
investigated the maximum available energy for gravitational wave
emission due to the micro-collapse triggered by the hadron-quark
phase transition in neutron stars. Moreover, the frequency and
damping time of the first axial $w$-mode of gravitational waves
have been studied for both hadronic and hybrid NSs. Since the most
uncertain part of the EOS of a neutron star is the density
dependence of the nuclear symmetry energy in the neutron-rich
nucleonic matter and the bag constant in the quark matter within
the MIT bag model, we have studied effects of the symmetry energy
and bag constant on the energy release as well as the frequency
and damping time of the first axial $w$-mode of GWs from neutron
stars. We have found that the energy release is much more
sensitive to the bag constant than the density dependence of the
nuclear symmetry energy. The frequency of the $w$-mode has been
found to be significantly different with or without the
hadron-quark phase transition and depends strongly on the bag
constant. Effects of the symmetry energy and bag constant on the
damping time have also been found to be appreciable but not as
strong as those on the frequency. We have further found that the
effect of the symmetry energy on the frequency becomes stronger
with a larger value of the bag constant that leads to a higher
hadron-quark transition density. While the predicted frequency of
the $w$-mode is significantly above the bands of operating
frequencies of the existing GW detectors, our results have
indicated that the frequency of the $w$-mode can indeed carry
important information about the internal structure of NSs and the
properties of dense neutron-rich matter.

\section{Acknowledgement}
This work was supported in part by the US National Science
Foundation Grant Nos. PHY-0757839 and PHY-0758115, the Welch
Foundation under Grant No.\ A-1358, the Texas Coordinating Board
of Higher Education Grant No.003565-0004-2007, the Young Teachers'
Training Program from China Scholarship Council under Grant No.
2007109651, the National Natural Science Foundation of China under
Grant No.10947023 and the Fundamental Research Funds for the
Central University, China under Grant No.2009ZM0193.

\end{document}